\documentclass[prd,preprint,superscriptaddress,showpacs,amsmath,amssymb]{revtex4}
\usepackage{graphicx}
\usepackage{dcolumn}
\usepackage{bm}

\begin{document}

\title{Effect of the Inverse Volume Modification in Loop Quantum Cosmology}
\author{Hua-Hui Xiong}
  \email{jimhard@163.com}
  \affiliation{Department of Physics, Beijing Normal University, Beijing 100875, China}

\author{Jian-Yang Zhu}
\thanks{Author to whom correspondence should be addressed}
  \email{zhujy@bnu.edu.cn}
  \affiliation{Department of Physics, Beijing Normal University, Beijing 100875, China}
\date{\today}

\begin{abstract}
It is known that in loop quantum cosmology (LQC) the universe avoids the
singularity by a bounce when the matter density approaches the critical
density $\rho _c$ (the order of Planck density). After incorporating the
inverse volume modifications both in the gravitational and matter part in
the improved framework of LQC, we find that the inverse volume modification
can decrease the bouncing energy scale, and the presence of nonsingular
bounce is generic. For the backward evolution in the expanding branch, in
terms of different initial states the evolution trajectories classify into
two classes. One class with larger initial energy density leads to the
occurrence of bounce in the region $a>a_{ch}$ where $a_{ch}$ marks the
different inverse volume modification region. The other class with smaller
initial energy density evolves back into the region $a<a_{ch}$. In this
region, both the energy density for the scalar field and the bouncing energy
scale decrease with the backward evolution. However, in the deep
modification region, because of the inverse volume modification the scalar
field is frozen, such that the bounce is present when the bouncing energy
scale decreases to be equal to the energy density of the scalar field. Using
numerical method, we show the evolution picture for the second class bounce.
\end{abstract}


\maketitle

\section{ Introduction}

In cosmology, an outstanding problem is the big bang singularity which is
expected to be solved by quantum gravity. Loop quantum cosmology (LQC) uses
the framework developed from loop quantum gravity (LQG) \cite{LQG} to deal
with the issues in cosmology \cite{lqc,mathematical-structure}. As a
ramification of LQG, LQC inherits the nonperturbative and background
independent feature of LQG. In LQC, the underlying geometry is discrete as
in LQG, and the universe evolution is described by the difference equation
which can go through the singularity nonsingularly \cite{singularity-free}.

The central technical issue in LQC is to quantize the Hamiltonian
constraint which generates the dynamical law. The classical
Hamiltonian constraint consists of the basic variables, i.e., the
SU(2) valued holonomies and the densitized triads. Conventionally,
on quantization of the Hamiltonian constraint, the SU(2) valued
holonomies in the gravitational part traces over the $J=1/2$
fundamental representation, while in the matter part the SU(2)
representation can be freely specified to define the inverse volume
resulting in the quantization ambiguity known as inverse scale
factor modification. Based on such quantum Hamiltonian constraint,
the effective dynamics can be obtained by some approximation. By use
of the effective dynamics one can investigate some phenomena in the
semiclassical region, where the spacetime recovers the continuum and
the difference equation is replaced by a differential equation
\cite{Date,semiclassical-state}. A number of results, such as a
natural inflation from quantum geometry \cite
{inflation-geometry,Xiong-Zhu}, avoidance of a big crunch in closed
cosmology \cite {avoidance-bigcrunch}, an oscillating universe with
suitable initial condition for inflation
\cite{oscillating-universe}, appearance of a cyclic universe
\cite{cyclic-universe} and a mass threshold of black hole \cite
{mass-threshold}, etc, are interesting and remarkable.

As shown in Ref. \cite{inverse-volume}, the quantization ambiguity also can
appear in the gravitational part. Using arbitrary $J$ representation for the
holonomies, the Hamiltonian constraint operator is constructed in Ref. \cite
{inverse-volume}. It shows that in the semiclassical region the
gravitational Hamiltonian gets modification similar to the inverse volume
modification in the matter part. This is because that, in the classical
Hamiltonian constraint, both the gravitational part and matter part contain
inverse volume terms, and, for quantization, promoting the inverse volume
into Poisson bracket between the volume and the holonomies leads to the same
quantization ambiguity \cite{quantization-ambiguity}. Therefore, as a
phenomenal investigation, in the semiclassical region the inverse volume
modification appearing in the gravitational part should also be included.

Recently, the semiclassical state is constructed in Refs. \cite
{bigbang,improved-dynamics}, and the most important result is that, by
evolving the semiclassical state backwards in the high energy density
regime, the expanding universe bounces into a contracting branch, so that
the singularity can be avoided. An effective Hamiltonian constraint
incorporating the discrete quantum geometry can well describe the evolution
of the semiclassical state \cite{improved-dynamics,preprint}. This
constraint predicts a quadratic density correction in the modified Friedmann
equation for the Hubble rate $H=\dot{a}/a$, $H^2\varpropto \rho (1-\rho
/\rho _c)$. The modified Friedmann equation implies a bounce when the matter
density approaches the critical density $\rho _c$ ($\rho _c$ is about $0.82$
times the Planck density). Using the modified Friedmann equation, some
interesting results are obtained: a bounce can happen avoiding of the
singularity when the energy density approaches the critical value $\rho _c$
\cite{nonsingular-bounce}; the scaling solutions of the modified Friedmann
equation have dual relationship with those in Randall-Sundrum cosmology \cite
{dual}; the future singularity can be avoided by the modified Friedmann
equation \cite{future-singularity}. However, in these works the inverse
volume modification is neglected both in the gravitational part and the
matter part, and for the gravitational part the holonomies are valued only
on the $J=1/2$ fundamental representation. Furthermore, from the viewpoint
of lattice model, a large spin $J$ means a longer range interaction between
lattices which appear as background structures \cite{lattice}, and the
inverse volume modification also could contribute to the seed of structure
formation \cite{structure}. The effect of the inverse volume modifications
to the effective dynamics is still unclear for the improved framework. Some
further questions still need to be answered: What is the effect of the
inverse volume modification in the gravitational part of the effective
dynamics? What is its effect on the matter part? We shall look into these
questions in the framework of the effective Hamiltonian.

In Ref. \cite{improved-dynamics}, the improved Hamiltonian constraint
operator is introduced. The minimal area gap can be better exploited by the
improved Hamiltonian to realize the physical idea, i.e., in the flat model a
bouncing universe is present purely because of the discrete quantum
geometry. In this paper, we work in the improved framework to analyze the
effect of the inverse volume modification both in the gravitational part and
matter part of the effective dynamics. We find that the inverse volume
modification in the gravity sector raised by a bigger quantization ambiguity
parameter $J$ could decrease the energy scale at the bounce point. And,
whether a bounce really happens or not depends on whether the matter density
at the high energy region attains the bouncing energy scale. The inverse
volume modification greatly changes the matter density below the scale $a_{*}
$. With the backward evolution in an expanding universe, in the region $%
a<a_{ch}$ the scalar field is frozen such that a nonsingular bounce is
present. Our numerical results show the genericness of bounce in LQC.

This article is organized as follows. In Sec. \ref{Sec. 2}, we give a brief
review over the Hamiltonian constraint and the modified Friedmann equation.
Then in Sec. \ref{Sec. 3}, the effect of the inverse volume modification in
universe evolution is analyzed in detail. Section \ref{Sec. 4} contains our
numerical results, displaying the evolution picture of the nonsingular
bouncing universe. Finally, the concluding remarks are made in Sec. \ref
{Sec. 5}.

\section{Hamiltonian constraint and modified Friedmann equation}

\label{Sec. 2}

\subsection{Classical and quantum Hamiltonian constraint}

LQC is a canonical quantization of cosmology model mimicing the quantizing
procedures as in the full theory. Imposing on the homogeneous and isotropic
symmetry in LQC, the classical phase space is reduced to two degrees of
freedom consisting of the conjugate connection $c$ and triad $p$. The
Poisson bracket between the conjugate variables satisfies $\left\{
c,p\right\} =\frac 13\gamma \kappa $, where $\kappa =8\pi G$ ($G$ is the
gravitational constant), and $\gamma $ is the dimensionless Barbero-Immirzi
parameter whose value, $\gamma \approx 0.2375$, set by the black hole
entropy calculation. For the flat model, the usual Friedman-Robertson-Walker
(FRW) metric expression can be identified with the relation given by
\begin{equation}
c=\gamma \dot{a},\ \left| p\right| =a^2,  \label{a1}
\end{equation}
where $a$ is the FRW scale factor, and the absolute value of $p$ denotes the
two orientations of the triad. Here, we only take the positive orientation.
In terms of the connection and triad, the classical Hamiltonian constraint
is given by \cite{mathematical-structure}
\begin{equation}
H=H_G+H_M,\ H_G=-\frac 3{\kappa \gamma ^2}\text{sgn}(p)\sqrt{\left| p\right|
}c^2.  \label{a2}
\end{equation}

On quantization in LQC, there is no operator directly corresponding to the
connection $c$. Instead, the elementary variables are the triad $p$ and the
holonomies $h_i(\mu )$ defined as the connection $c$ along an edge, i.e., $%
h_i(\mu )=e^{\mu c\tau _i}$, where $\mu $ is the length of the $i$th edge,
and $\tau _i$ is a basis in the Lie algebra su(2) satisfying $\left[ \tau
_i,\tau _j\right] =\varepsilon _{ijk}\tau ^k$. As in LQG, the holonomies and
triads have well-defined operators. So, the Hamiltonian constraint given by
Eq. (\ref{a2}) must be reformulated in terms of the holonomies and the
triads. The gravitational part of the Hamiltonian constraint $H_G$ in the
full theory can be written as \cite{mathematical-structure}
\begin{equation}
H_G=-\frac 1{\kappa \gamma ^2}\int d^3x\varepsilon _k^{ij}\frac{E_i^aE_j^b}{%
\sqrt{\left| \det E\right| }}F_{ab}^k,  \label{a3}
\end{equation}
where $F_{ab}^k$ is the curvature component of the connection.

The strategies for quantization are: first, to reformulate the Hamiltonian
constraint by the holonomies and triads, and then, to promote these
elementary variables to the corresponding operators. The term $E_i^aE_j^b/%
\sqrt{\left| \det E\right| }$ contains the inverse volume, and on
quantization the inverse volume is promoted to the commutator of the
holonomy and volume, resulting in the quantization ambiguity labeled by the
spin representation $J$. For the $J=1/2$ fundamental representation, the
quantum operator for the gravitational part is given by
\begin{eqnarray}
\widehat{H}_G &=&\frac i{4\pi \kappa \ell _p^2\gamma ^3\overline{\mu }^3}%
\sum_{ijk}\varepsilon ^{ijk}Tr\left( \hat{h}_i\hat{h}_j\hat{h}_i^{-1}\hat{h}%
_j^{-1}\hat{h}_k\left[ \hat{h}_k^{-1},\hat{V}\right] \right)  \nonumber \\
&=&\frac{6i}{\pi \kappa \ell _p^2\gamma ^3\bar{\mu}^3}\sin ^2\left( \frac{%
\bar{\mu}c}2\right) \cos ^2\left( \frac{\bar{\mu}c}2\right)  \nonumber \\
&&\times \left( \sin \left( \frac{\bar{\mu}c}2\right) \hat{V}\cos \left(
\frac{\bar{\mu}c}2\right) -\cos \left( \frac{\bar{\mu}c}2\right) \hat{V}\sin
\left( \frac{\bar{\mu}c}2\right) \right) ,  \label{a4}
\end{eqnarray}
where $\ell _p^2=G\hbar $ ($\hbar $ is the reduced Planck constant) and the
volume operator $\hat{V}=\hat{p}^{3/2}$. In Eq. (\ref{a4}) the holonomies
are valued along the edges of the square loop whose area is fixed by the
minimal eigenvalue of area operator in LQG. The physical area of the square
loop is $\bar{\mu}^2p=\alpha \ell _p^2$, where $\alpha $ is order one. The
area of the loop fixed by the minimal area of LQG shows the discrete feature
of quantum geometry. We shall see that this point essentially captures the
effect of quantum geometry making the effective dynamics very different from
the classical case in the high energy regime. As for the arbitrary $J$
representation, the gravitational Hamiltonian operator can be constructed as
done in Ref. \cite{inverse-volume}. One only replaces the fixed parameter $%
\mu _0$ with $\bar{\mu}$. As shown in Ref. \cite{improved-dynamics},
replacing $\mu _0$ with $\bar{\mu}$ is a key step to realize the improved
Hamiltonian constraint. After obtaining the gravitational Hamiltonian
constraint operator for arbitrary $J$ representation, it is useful to
extract the effective theory which is expected to capture the key feature of
the Hamiltonian constraint operator in the semiclassical region. Next, we
will show the inverse volume modification raised by the quantization
ambiguity parameter $J$ in the effective theory.

\subsection{Effective Hamiltonian constraint}

It is shown in Ref. \cite{improved-dynamics} that the quantum evolution
determined by the quantum Hamiltonian constraint is nonsingular in big bang
point. By evolving the semiclassical states backwards, the universe
evolution indicates that the big bang is replaced by a bounce when the
matter density approaches the critical value $\rho _c$. This feature can be
well described by the effective Hamiltonian constraint in the region which
is above the Planck scale and where spacetime recovers the continuum. The
effective Hamiltonian can be obtained by the WKB approximation \cite
{Date,semiclassical-state} or by a systematic way \cite{low-energy}. Here,
the effective Hamiltonian for the arbitrary $J$ representation is given by
\cite{dual,low-energy,vandersloot}
\begin{equation}
H_{eff}=-\frac 3{\kappa \gamma ^2\bar{\mu}^2}S_J\left( p\right) \sin
^2\left( \bar{\mu}c\right) +H_M,  \label{a5}
\end{equation}
where $S_J\left( p\right) $ encodes the inverse volume modification for the
gravitational part and is given by
\begin{equation}
S_J\left( p\right) =\sqrt{p}S\left( \frac p{p_J}\right) ,  \label{a6}
\end{equation}
and
\begin{eqnarray}
S\left( q\right) &=&\frac 4{\sqrt{q}}\left\{ \frac 1{10}\left[ \left(
q+1\right) ^{\frac 52}+sgn\left( q-1\right) \left| q-1\right| ^{\frac 52%
}\right] \right.  \label{a7} \\
&&\left. -\frac 1{35}\left[ \left( q+1\right) ^{\frac 72}-\left|
q-1\right| ^{\frac 72}\right] \right\},  \nonumber
\end{eqnarray}
where $p_J=\frac{8\pi \gamma J\bar{\mu}}3\ell _P^2$ mark the modification
region, below which (i.e., $q<1$) the modification is dramatic, and above
which the effect of the inverse volume modification can be neglected with $%
S\left( q\right) \approx 1$.

\subsection{Modified Friedmann equation}

\subsubsection{Small $J$ modification}

One can assume that $J$ is small, then $S_J\left( p\right) \approx \sqrt{p}$
and the effective Hamiltonian becomes
\begin{equation}
H_{eff}=-\frac 3{\kappa \gamma ^2\bar{\mu}^2}\sqrt{p}\sin ^2\left( \bar{\mu}%
c\right) +H_M.  \label{a8}
\end{equation}
Such an effective Hamiltonian presents a singularity-free bouncing universe
with the modified Friedmann equation \cite
{improved-dynamics,nonsingular-bounce}
\begin{equation}
H^2=\frac \kappa 3\rho \left( 1-\frac \rho {\rho _c}\right),
\label{a9}
\end{equation}
where $H$ is the Hubble parameter, the critical density $\rho _c=3/\kappa
\gamma ^2\bar{\mu}^2a^2=3/\kappa \gamma ^2\alpha \ell _p^2$ and the matter
density $\rho =H_M/p^{3/2}$. Based on the effective equation (\ref{a9}) some
interesting issues have been discussed \cite
{nonsingular-bounce,dual,future-singularity}.

\subsubsection{Large $J$ modification}

Now, we discuss the large $J$ modification to the effective dynamics. By the
effective Hamiltonian constraint given by Eq. (\ref{a5}), the Hamiltonian
equation can be obtained as
\begin{equation}
\dot{p}=\left\{ p,H_{eff}\right\} =-\frac{\gamma \kappa }3\frac{\partial
H_{eff}}{\partial c}.  \label{b0}
\end{equation}
The Hamiltonian equation (\ref{b0}) gives out
\begin{equation}
a\dot{a}=\frac 1{\gamma \bar{\mu}}S_J\left( a^2\right) \sin \left( \bar{\mu}%
c\right) \cos \left( \bar{\mu}c\right) .  \label{b1}
\end{equation}
Squaring the above equation and then making use of the Hamiltonian
constraint $H_{eff}\approx 0$, the modified Friedmann equation incorporating
the inverse volume modification can be obtained as
\begin{equation}
H^2=\frac \kappa 3\rho \left( S\left( q\right) -\frac \rho {\rho
_c}\right), \label{b2}
\end{equation}
where $q\equiv p/p_J=\left( a/a_{g*}\right) ^2$.

\section{Effect of the inverse volume modification in universe evolution}

\label{Sec. 3}

\subsection{Modification in the gravitational sector}

The modified Friedmann Eq. (\ref{b2}) gives out the evolution of the
universe. In the early time of the universe, if the matter density $\rho
=S\left( q\right) \rho _c$, then there exists a turn-around. At the
turn-around, a bounce or a collapse happens depending on the sign of the
second derivative of the triad with time. For the effective theory one can
analyze the bounce behavior of the universe as discussed in Ref. \cite
{nonsingular-bounce}. From the Hamiltonian Eq. (\ref{b0}) and $\dot{c}%
=\left\{ c,H_{eff}\right\} $, one can get
\begin{eqnarray}
\ddot{p}\left| _{\dot{p}=0}\right.  &=&\kappa S_J^2(p)\left[ \frac 23\left(
1+\frac{d\ln S_J\left( p\right) }{d\ln p}\right) \rho _c-\frac 1{S\left(
q\right) }\left( \frac 23\frac 1{\sqrt{p}}\frac{\partial H_M}{\partial p}%
\right) \right]   \nonumber \\
&=&\kappa S_J^2(p)\left[ \frac 23\left( 1+\frac{d\ln S_J\left( p\right) }{%
d\ln p}\right) \rho _c+\frac 1{S\left( q\right) }P\right] ,  \label{b4}
\end{eqnarray}
where the second line uses the definition of the pressure $P\equiv \frac 23%
\frac 1{\sqrt{p}}\frac{\partial H_M}{\partial p}$ as in Ref. \cite
{pressure-definition}. For a constant state parameter equation, i.e., $%
P=\omega \rho $, the Eq. (\ref{b4}) can be written as
\begin{equation}
\ddot{p}\left| _{\dot{p}=0}\right. =\kappa S_J^2(p)\rho _c\left[ \frac 23%
\left( 1+\frac{d\ln S_J\left( p\right) }{d\ln p}\right) +\omega \right] .
\label{b5}
\end{equation}
For $p<p_J$, the function $d\ln S_J\left( p\right) /d\ln p$ monotonically
decreases, and when $a\ll a_{*}=\sqrt{p_J}$, $\ddot{p}\left| _{\dot{p}%
=0}\right. \rightarrow \kappa S_J^2(p)\rho _c\left( \frac 43+\omega \right) $%
; $a\rightarrow a_{*}$, $\ddot{p}\left| _{\dot{p}=0}\right. \rightarrow
\kappa p_J\rho _c\left( 1.05+\omega \right) $. Therefore, one can obtain
that the recollapse requires $\omega <-1$, which classically violates the
null energy condition and implies an expanding universe with increasing
energy density. This exotic matter behaves like phantom field. In the next
section, we shall explicitly give out the expression of the effective state
parameter equation $\omega _{eff}$; it will show that the presence of a
turn-around always leads to a bouncing universe. So, for the universe with
quantum modification, the state parameter equation satisfies $\omega >-1$.

In order to better understand the effect of the inverse volume
modification on the modified Friedmann, we shall distinguish the two
basic scales: One is the bounce scale $a_{bounce}$, which is
determined by the condition $\rho \left( a\right) =S\left( q\right)
\rho _c$, and whose explicit value is related with the mater content
(in the next section we will deal with the related issue). The other
one is the inverse volume modification scale
$a_{*}=\sqrt{p_J}=\sqrt{8\pi \gamma J\bar{\mu}/3}\ell
_P=\left( 8\pi \alpha ^{1/2}\gamma /3\right) ^{1/3}J^{1/3}\ell _P$ ($\bar{\mu%
}$ is replaced by the fixed area relation $\bar{\mu}^2a^2=\alpha \ell _P^2$%
), whose magnitude depends on the parameter $J$. We know that $a_{bounce}$
is the minimal scale for the evolution of the universe. If the inverse
volume modification scale $a_{*}<a_{bounce}$, the effect of the inverse
volume modification should be neglected because in this range the inverse
volume modification tends to its classical form $S_J\left( p\right) =\sqrt{p}
$. However, for a large enough $J$, such that $a_{bounce}<a<a_{*}$, then the
inverse volume modification becomes notable in LQC.

From the above analysis, we know that for the bouncing universe the matter
density at the bouncing energy scale is $\rho _{bounce}=S\left( q\right)
\rho _c$. In the region $a_{bounce}<a<a_{*}$, the function $S\left( q\right)
$ monotonically increases and $S\left( q\right) <1$. For $a_{bounce}\ll
a_{*} $, $S\left( q\right) \approx \frac 65\frac a{a_{*}}\ll 1$. Therefore,
we can conclude that for LQC the inverse volume modification in the
gravitational part reduce the matter density at the bounce point.

In this subsection, we have analyzed the effect of the inverse volume
modification to the bouncing universe. However, for LQC, whether a bounce
happens or not depends on the content of the matter field. In the next
section, we will show the effect of the inverse volume modification in the
matter sector on the evolution of LQC.

\subsection{Modification in the matter sector}

In this section, we focus on the inverse volume modification in the matter
sector for the modified Friedmann equation. In earlier literatures, the
SU(2) representation for the matter part is labeled by a half integer which
can be freely specified independent of the $J$ representation in the
gravitational part. As argued in Ref. \cite{inverse-volume}, it is more
natural to take the same representation for them. Therefore, for a massive
scalar field with self-interaction potential $V\left( \phi \right) $, the
Hamiltonian $H_M$ is given by \cite{inflation-geometry}
\begin{equation}
H_M=\frac 12d_J\left( a\right) p_\phi ^2+a^3V\left( \phi \right) ,
\label{c1}
\end{equation}
where $p_\phi $ is the conjugate momentum for the scalar field $\phi $, and $%
d_J\left( a\right) $ is the eigenvalue of the inverse volume operator $%
\hat{a}^{-3}$ which can be approximately described as
\begin{equation}
d_J\left( a\right) =D\left( \frac{a^2}{a_{*}^2}\right) a^{-3}.  \label{c2}
\end{equation}
In which, the modification region is still marked by the scale $a_{*}=\left(
8\pi \alpha ^{1/2}\gamma /3\right) ^{1/3}J^{1/3}\ell _P$, and the function $%
D\left( a^2/a_{*}^2\right) $ respects the inverse volume modification and
has the form
\begin{eqnarray}
D\left( q\right)  &=&\left( \frac 8{77}\right) ^6q^{\frac 32}\left\{ 7\left[
\left( q+1\right) ^{\frac{11}4}-\left| q-1\right| ^{\frac{11}4}\right]
\right.   \nonumber \\
&&-\left. 11q\left[ \left( q+1\right) ^{\frac 74}-sgn\left( q-1\right)
\left| q-1\right| ^{\frac 74}\right] \right\} ^6.  \label{c3}
\end{eqnarray}
Above the scale $a_{*}$, $D\left( q\right) \simeq 1$, so the inverse volume
modification can be neglected. While at $q\ll 1$, $D\left(
a^2/a_{*}^2\right) \approx \left( 12/7\right) ^6\left( a/a_{*}\right) ^{15}$.

Now, the modified Friedmann equation becomes
\begin{equation}
H^2=\frac \kappa 3\rho \left[ S\left( a^2/a_{*}^2\right) -\frac \rho {\rho _c%
}\right]  \label{c4}
\end{equation}
where the matter density $\rho =H_M/p^{3/2}$ which receives the inverse
volume modification.

From the Hamiltonian equation $\dot{\phi}=\left\{ \phi ,H_M\right\} $, one
can get
\begin{equation}
p_\phi =d_J^{-1}\left( a\right) \dot{\phi}.  \label{c5}
\end{equation}
It leads to the matter density
\begin{equation}
\rho =\frac 12D^{-1}\dot{\phi}^2+V\left( \phi \right) .  \label{c5-1}
\end{equation}

Differentiating Eq.(\ref{c5}) and substituting it into another Hamiltonian
equation $\dot{p}_\phi =\left\{ p_\phi ,H_M\right\} $, the modified
Klein-Gordon equation can be obtained as
\begin{equation}
\ddot{\phi}=-3H\left( 1-\frac 13\frac{d\ln D}{d\ln a}\right) \dot{\phi}%
-DV^{^{\prime }}\left( \phi \right) ,  \label{c6}
\end{equation}
where prime denotes $\frac d{d\phi }$. Then the time derivative of the
matter density is
\begin{equation}
\dot{\rho}=-3HD^{-1}\dot{\phi}^2\left( 1-\frac 16\frac{d\ln D}{d\ln a}%
\right) ,  \label{c6-1}
\end{equation}
where $d\ln D/d\ln a$ is a monotonically decreasing function in the region $%
a<a_{*}$. So, there exists a characteristic scale $a_{ch}\approx 0.91a_{*}$
determined by $d\ln D/d\ln a=6$, and the mater density attains the maximal
value. Classically, with the expansion of the universe, the matter density
decreases. However, in the region of $a<a_{ch}$, the matter density
increases; above this region the matter density starts to decrease and
enters the regime where the quadratic density modification dominates and the
effect of the inverse volume modification can be neglected. In the
following, we will discuss the possible evolution of LQC in different
regions.

First, it is useful to identify the effective matter density and pressure.
From Eq. (\ref{c4}), the modified Raychoudhuri equation is written as
\begin{eqnarray}
\frac{\ddot{a}}a &=&\dot{H}+H^2  \nonumber \\
&=&-\frac \kappa 6\left\{ -2\rho \left( S-\frac \rho {\rho _c}\right) -\frac %
1H\left( \dot{\rho}S+\rho \dot{S}-\frac{2\rho \dot{\rho}}{\rho _c}\right)
\right\} .  \label{c7}
\end{eqnarray}
Then the effective energy density and pressure can be defined as
\begin{equation}
\rho _{eff}=\rho \left( S-\frac \rho {\rho _c}\right) ,  \label{c8}
\end{equation}
\begin{equation}
P_{eff}=-\rho \left( S-\frac \rho {\rho _c}\right) +\frac 1{3H}\left( \dot{%
\rho}S+\rho \dot{S}-\frac{2\rho \dot{\rho}}{\rho _c}\right) .  \label{c9}
\end{equation}
They satisfy the conservative equation
\begin{equation}
\dot{\rho}_{eff}+3H\left( \rho _{eff}+P_{eff}\right) =0.  \label{c10}
\end{equation}

Substituting Eq. (\ref{c6-1}) into the expression of $P_{eff}$, the
effective state parameter equation is
\begin{equation}
\omega _{eff}=\frac{P_{eff}}{\rho _{eff}}=-1+\frac{2\dot{\phi}^2\left( 1-%
\frac 16\frac{d\ln D}{d\ln a}\right) }{\dot{\phi}^2+2DV\left( \phi \right) }%
\frac{S-\frac{2\rho }{\rho _c}}{S-\frac \rho {\rho _c}}-\frac 23q\frac{\frac{%
dS}{dq}}{S-\frac \rho {\rho _c}},  \label{c11}
\end{equation}
where the function $dS/dq$ monotonically decreases in the region $a<a_{*}$,
and asymptotically vanishes beyond this region. From the effective parameter
equation we know that both the inverse volume modification and the quadratic
density correction can contribute to the inflation of the universe.

In LQC, the evolution trajectory is determined by the dynamical initial
condition \cite{dynamical-condition}. So, we evolve the universe at large
scale $a\gg $ $a_{*}$ backwards in the expanding universe where the matter
density $\rho \ll \rho _c$ and the quantum modifications can be neglected.
With the backward evolution, the matter density increases in the region $%
a>a_{ch}$. If the matter density satisfies $\rho \geqslant S\left(
a_{ch}^2/a_{*}^2\right) \rho _c$, then $\rho _{eff}=0$ leads to a bouncing
universe which is caused purely by the quadratic density modification. This
situation is similar to the case of $a_{bounce}\geqslant a_{*}$, where the
inverse volume modification can be neglected. If the mater density $\rho
<S\left( a_{ch}^2/a_{*}^2\right) \rho _c$, the universe evolves back to the
deep inverse volume modification region $a<a_{ch}$. In the region $a<a_{ch}$%
, from Eqs. (\ref{a7}) and (\ref{c6-1}) we know that both the matter density
$\rho $ and the bouncing energy scale $S(q)\rho _c$ decrease with the
backward evolution. If the bouncing energy scale $S(q)\rho _c$ decreases
faster than the matter density, then there is no bounce; the universe will
evolve towards the big bang point where the effective theory is invalid and
the quantum difference equation takes over the evolution of the universe.
Another possible way is that the mater density decreases more slowly than
the bouncing energy scale, until $\rho =S(q)\rho _c$, where a bounce happens
and the universe is bounced into a contracting branch. We would like to
point out that the actual evolution way favors the latter. It is because
that, with the backward evolution, the density decrease. However, for the
density $\rho =\frac 12D^{-1}\dot{\phi}^2+V\left( \phi \right) $, in the
small volume, the term $D^{-1}\varpropto \left( a/a_{*}\right) ^{15}$ is far
larger than one. This needs the velocity $\dot{\phi}$ to be extremely small,
such that the field is frozen. So, the kinetic term can be neglected, and
the potential dominates. With the decreasing scale $a$, the field is frozen,
until the bouncing energy scale $S\left( a^2/a_{*}^2\right) \rho _c$
decreases to be equal to the potential $V\left( \phi \right) $. At that
point, a bounce happens. In what follows, using the numerical way, we will
show the evolution of the universe.

\section{Genericness of bounce shown in numerical results}

\label{Sec. 4} In the inverse volume modification region, a negative scalar
potential can lead to a bouncing universe only when the condition $\rho =0$
is satisfied\cite{cyclic-universe,early-universe}. This is also true for the
modified Friedmann equation presented in this paper. From the modified
Raychoudhuri Eq. (\ref{c7}), it is easily seen that in the region $a<a_{ch}$
the vanishing density implies a bounce. Such bounce purely comes from the
inverse volume modification. The reason is that the inverse volume
modification in this region makes
\[
\frac{\ddot{a}}a\left| _{\dot{a}=0}\right. =-\frac \kappa 2D^{-1}\dot{\phi}%
^2\left( 1-\frac 16\frac{d\ln D}{d\ln a}\right) S\left( q\right) >0,
\]
which implies a bounce. In the following, we will show the evolution
trajectory of LQC with a positive potential.

Using the modified Friedmann Eq. (\ref{c4}) and the Klein-Gordon Eq. (\ref
{c6}), we can obtain the dynamical law for the universe with a massive
scalar field with the potential $V\left( \phi \right) =m^2\phi ^2/2$. Here,
for the numerical calculation we take the mass $m=0.2$. From the above
discussion we know that these trajectories can be divided into two classes:
the ones with higher energy density can be bounced into an contracting
branch in the region $a>a_{ch}$, and the others with lower energy density $%
\rho <S\left( a_{ch}^2/a_{*}^2\right) \rho _c\approx 0.94\rho _c$
evolve back to the deep modification region $a<a_{ch}$. In this
paper, we are only concerned about the latter. For the backward
evolution, we take the initial state marked by the energy density
$\rho _{in}$ at the scale $a_{ch}$. Our numerical results show that
the presence of a bounce is generic for theses trajectories with the
initial state $\rho _{in}$ in the region $\left( 0.001\sim 1\right)
S\left( a_{ch}^2/a_{*}^2\right) \rho _c$. Below this region, we
think that it is not suitable to take the energy density far below
the Planck density as the initial state in the quantum modification
region. The initial state can be written as
\begin{eqnarray}
\rho _{in}\left( a_{ch}\right)  &=&\frac 12D^{-1}\left(
a_{ch}^2/a_{*}^2\right) \dot{\phi}^2+\frac 12m^2\phi ^2  \nonumber \\
&=&\lambda S\left( a_{ch}^2/a_{*}^2\right) \rho _c,  \label{d1}
\end{eqnarray}
where $\lambda $ denotes the initial state with different initial energy.
From Eqs. (\ref{c4}) and (\ref{c6}) we know that under the rescaled
transformation $a\rightarrow a/a_{*}$ the dynamical equations do not change
and do not depend on the scale $a$ explicitly. So, under this rescaling we
can take an arbitrary value of $J$ to show the evolution trajectories. For
other choices of $J$, the evolution trajectories are the same as the chosen
one. Here, we take $J=1000$ to show the evolution picture.

\begin{figure}[tbp]
\includegraphics[clip,width=0.6\textwidth]{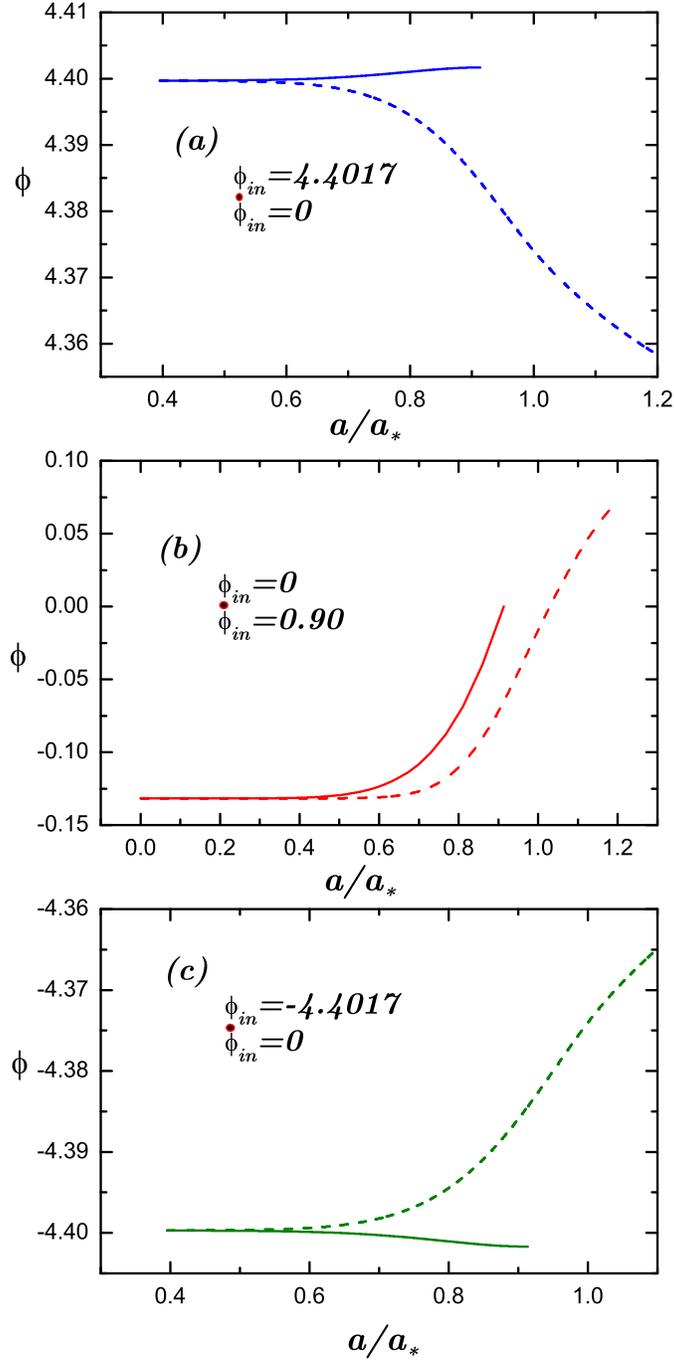}
\caption{The evolution of the scale factor is shown for a scalar
field with potential $V\left( \phi \right) =m^2\phi ^2/2$ for
different initial states distinguished by different initial values
of $\phi _{in}$(these initial states are constrained by the fixed
initial energy density $\rho _{in}=0.5S\left(
a_{ch}^2/a_{*}^2\right) \rho _c\approx 0.74\rho _c$ at the scale
$a_{ch}$, and $\dot{\phi}_{in}$ is determined by Eq.\ref{d1}). The
solid lines denote the expanding branch, and the dashed lines show
the contracting branch. In the region $a<a_{ch}\approx 0.91a_{*}$,
the backward evolution shows that for bigger initial value of
$\left| \phi _{in}\right| $ the bounce happens at bigger scale.}
\label{Fig.1}
\end{figure}

\begin{figure}[tbp]
\includegraphics[clip,width=0.6\textwidth]{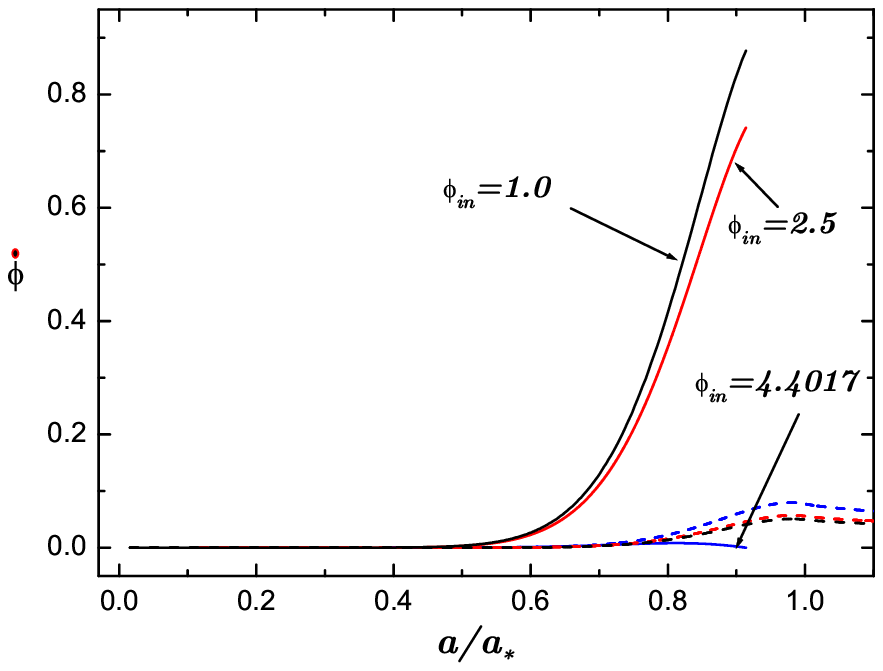}
\caption{The sketch shows that near the bouncing scale the scalar
field is frozen such that the kinetic term vanishes and the
potential dominates. Here, the solid lines and the dashed lines
denote, respectively, the expanding branch and the contracting
branch. The initial states take values as in Fig. 1 constrained by
the fixed initial energy density $\rho _{in}=0.5S\left(
a_{ch}^2/a_{*}^2\right) \rho _c\approx 0.74\rho _c$.} \label{Fig.2}
\end{figure}

Figure \ref{Fig.1} shows the presence of bounce for $\lambda =0.5$.
For different initial values of $\phi $, the bouncing scale is
different. With larger value of $\left| \phi \right| $, the bouncing
scale is larger. The contrary is that with larger value of $\left|
\dot{\phi}\right| $, the bounce happens at deeper region. This can
be regarded as that, for the backward evolution, a larger kinetic
term takes longer distance to be frozen. At the bouncing point, the
field is frozen at some fixed value of $\phi $. Figure \ref{Fig.2}
shows that with the backward evolution the scalar field is frozen
and then bounced into a contracting branch. For other choices of
$\lambda $, the bouncing behavior is similar as presented in Figs.
\ref{Fig.1} and \ref{Fig.2}. The difference is that with smaller
value of $\lambda $ the bounce happens at deeper region.

Furthermore, in the inverse volume modification region the scalar
field can be pushed to climb up the potential hill. As a result,
this can set the initial condition for a classical inflation
\cite{CMB-spectrum}. To achieve 60 e-folds of inflation for a
quadratic potential $V=m^2\phi ^2/2$ , the field must be displaced
by at least $3M_{Pl}$ from the minimum of its potential at the onset
of inflation. For the case of $\lambda =0.5$ presented in the above,
this means that for a successful inflation the bounce should happen
at bigger scale for larger value of $\left| \phi _{in}\right| $
shown in Fig. \ref{Fig.1}.

\section{Conclusion}

\label{Sec. 5} In this paper, we mainly discuss the effect of the
inverse volume modifications on the effective dynamics, which
implies a nonsingular bounce for the evolution of the universe when
the matter density approaches the high energy region. This is a
nontrivial result compared with the bouncing mechanism in the
fundamental $J=1/2$ representation \cite
{improved-dynamics,nonsingular-bounce}. Because until now there is
still lack of a complete quantization for large $J$ representation
in the improved framework, we expect that the evolution picture
presented in this paper will be given by a complete quantum theory
as in \cite{improved-dynamics}. Furthermore, we argue that the
genericness of bounce in the high spin representation is not
accidental: the presence of bounce is a consequence of the fact that
LQC essentially incorporates two quantum effects, i.e., the inverse
volume modifications and the quadratic density modification raised
by the minimal area gap, and the two quantum effects jointly induce
the occurrence of bounce. For the inverse volume modification, the
modification region is marked by the scale $a_{*}$, below which the
modifications are notable and above which the modifications become
weak. The inverse volume modification in the gravity sector
decreases the bouncing energy scale in the region
$a_{bounce}<a<a_{g*}$ (neglecting the inverse volume modification
the bouncing energy scale is equal to the critical density $\rho _c$
). And, the inverse volume modification in the matter sector changes
the classical behavior of density. Below the scale $a_{ch}\approx
0.91a_{*}$ the time derivative of the matter density is positive,
implying that, with the backward evolution of an expanding universe,
the matter density decreases. So, in the region $a<a_{ch}$ the
presence of bounce depends on the relative decreasing rate between
the matter density and the bouncing energy scale. However, in the
deep quantum modification region the scalar field is frozen. So the
kinetic term can be neglected, and the matter density is dominated
by the potential. Therefore, the matter density is almost frozen at
a fixed value. With the backward evolution, a bounce happens until
the bouncing energy scale decreases to be equal to the potential.
Our numerical results shown in Figs. \ref{Fig.1} and \ref{Fig.2}
provide a illustration of the evolution pictures. For different
initial values of the matter density the bouncing scale is
different.

In earlier literatures, the SU(2) representation for the gravitational
Hamiltonian is taken to be the fundamental representation, but for the
matter Hamiltonian the representation is freely specified. If one does so,
then in this paper the modified Friedmann equation becomes $H^2=\kappa \rho
\left( 1-\rho /\rho _c\right) /3$, where $\rho =\frac 12D^{-1}\dot{\phi}%
^2+V\left( \phi \right) $ receives the inverse volume modification. For the
backward evolution in the expanding branch, if the matter density increases
to be equal to the critical density $\rho _c$ in the region $a>a_{ch}$, a
bounce happens. If the universe evolves back through the scale $a_{ch}$, the
matter density starts to decrease, so a bounce will never be present for the
universe with a positive potential. This is a distinct difference between
the two views of the inverse volume modification. Moreover, the inverse
volume modification can leave indirect imprint on the CMB spectrum\cite
{CMB-spectrum}. We expect the further research could put constraint for the
value of $J$ based on the prediction for observations.

In this paper we analyze the effect of the inverse volume modifications on
the improved LQC, and we show that the presence of nonsingular bounce is
generic.

\acknowledgments
The work was supported by the National Basic
Research Program of China (2003CB716302).

\end{document}